\documentclass{sigchi-ext}
\usepackage[T1]{fontenc}
\usepackage{textcomp}
\usepackage[scaled=.92]{helvet} 
\usepackage{graphicx} 
\usepackage{balance}  
\usepackage{booktabs} 
\usepackage{ccicons}  
\usepackage{ragged2e} 
\usepackage{subfigure}
\usepackage{multicol}

\title{\textit{I Ate This}: A Photo-based Food Journaling System with Expert Feedback}

\numberofauthors{6}
\author{%
  \alignauthor{%
    \textbf{Shubham Goyal}\\
    \affaddr{Holmusk} \\
    \email{shubham.goyal@holmusk.com} }
    \alignauthor{%
    \textbf{Waqas Awan}\\
    \affaddr{Holmusk}\\
    \email{waqas.awan@holmusk.com} } \vfil 
    \alignauthor{%
    \textbf{Qi Liu}\\
    \affaddr{School of Computing, National University of Singapore}\\
    \email{qiliu@comp.nus.edu.sg} }
    \alignauthor{%
    \textbf{Bimlesh Wadhwa}\\
    \affaddr{School of Computing, National University of Singapore}\\
    \email{dcsbw@nus.edu.sg} } \vfil 
    \alignauthor{%
    \textbf{Khairina Tajul-Arifin}\\
    \affaddr{Holmusk}\\
    \email{ kay.tarifin@holmusk.com} }
    \alignauthor{%
    \textbf{Zhenguang Liu}
    \affaddr{National University of Singapore}
    \email{lzg@nus.edu.sg} }
    }
\def\plaintitle{SIGCHI Extended Abstracts Sample File: Note Initial
  Caps} \def\plainauthor{First Author, Second Author, Third Author,
  Fourth Author, Fifth Author, Sixth Author}
\def\plainkeywords{Authors' choice; of terms; separated; by
  semicolons; include commas, within terms only; required.}

\hypersetup{%
  pdftitle={\plaintitle}, pdfauthor={\plainauthor},
  pdfkeywords={\plainkeywords}, }

\begin{document}

\maketitle

\RaggedRight{} 

\begin{abstract}
What we eat is one of the most frequent and important health decisions we make in daily life, yet it remains notoriously difficult to capture and understand. Effective food journaling is thus a grand challenge in personal health informatics.
In this paper we describe a system for food journaling called \textit{I Ate This}, which is inspired by the Remote Food Photography Method (RFPM). \textit{I Ate This} is simple: you use a smartphone app to take a photo and give a very basic description of any food or beverage you are about to consume. Later, a qualified dietitian will evaluate your photo, giving you feedback on how you did and where you can improve. The aim of \textit{I Ate This} is to provide a convenient, visual and reliable way to help users learn from their eating habits and nudge them towards better choices each and every day. Ultimately, this incremental approach can lead to long-term behaviour change. Our goal is to bring RFPM to a wider audience, through APIs that can be incorporated into other apps.
\end{abstract}



\section{Introduction}
Food choices are among the most frequent and important health decisions in daily life. Recording and providing feedback on these choices can be effective in prompting people to be more mindful of the quality and quantity of foods they consume. This can be crucial to support a variety of health goals, including: weight loss, healthier eating, identifying allergies, detecting deficiencies and diabetes control. Smartphone apps have made food journaling and feedback accessible to many people. They allow users to search for components of a meal against a food database, enter portion information, and journal calories and other nutritional information relative to a daily goal. MyFitnessPal\footnote{https://www.myfitnesspal.com/} and FatSecret\footnote{https://www.fatsecret.com/} are among the most popular apps in this category. Many apps incorporate features such as custom recipes, shortcuts to commonly eaten foods and barcode scanning to make entry easier. This data allows apps to provide real-time, personalised feedback on performance to either reinforce current behaviour or highlight areas for improvement.

However, manual logging of foods and calories counting, even on smartphones, requires a high level of engagement and is burdensome for most people \cite{barrett:1991}\cite{craig:2000}. Another issue is inaccurate and misleading results owing to difficulty finding appropriate entries in food databases, unreliable entries in those databases and difficulty estimating portions \cite{Cordeiro:2015}. Studies suggest that journaled calories can be out by as much as $20-50\%$ \cite{lansky:2000}. These issues detract from the benefits and impact of food journaling and also result in poor adoption and adherence. One study found that only $3\%$ of $190,000$ downloads of a food-journal app resulted in a person using the mobile food journal for more than one week \cite{Helander:2014}.

\begin{figure}[t]
	\includegraphics[width=0.9\columnwidth]{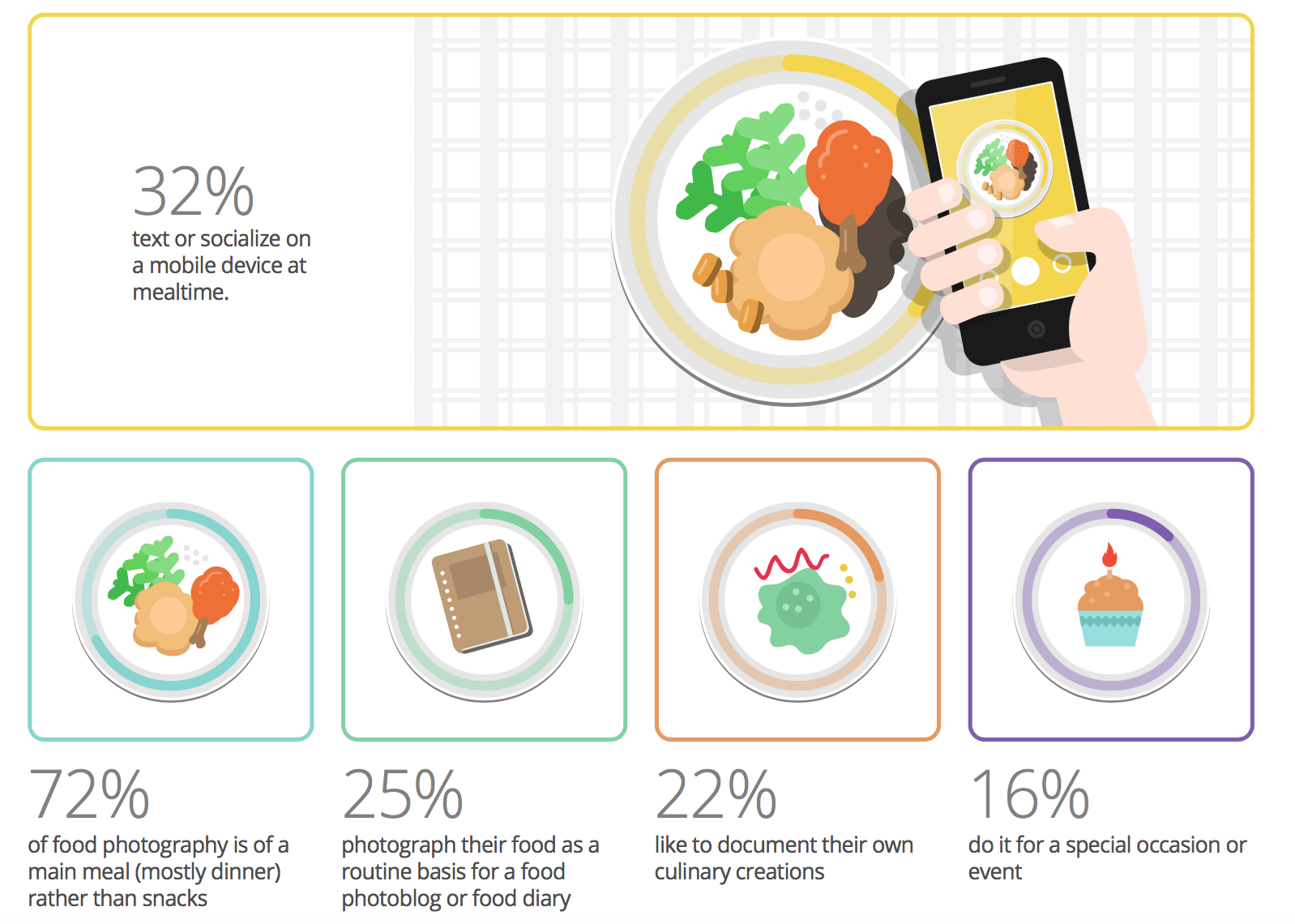}
	\caption{Food Photography and Social Media}~\label{fig:data}
\end{figure}

While food journaling is not making a huge societal impact, photographing meals and sharing the images on social media is a growing phenomenon \ref{fig:data}. Across sites like Twitter, Facebook, Instagram and Flickr, photos with food-related tags are among the most frequent and popular posts, and have grown exponentially in recent years. A new generation of apps such as Burple\footnote{http://blog.burpple.com/}, FoodSpotting\footnote{http://www.foodspotting.com/find/in/Singapore} and SnapDish\footnote{http://snapdish.co/} are specifically dedicated to photographing and sharing food. Photo-based food journaling allows for rapid and easy capture that lowers the barrier and levels the playing field for journaling any meal type \cite{Cordeiro:rethink:2015}. With traditional journaling methods, packaged or fast foods are easiest to log, whereas home cooked or and certain mixed dishes are more challenging.

Most existing photo-based food journaling approaches attempt to go a step beyond just capturing and presenting food images for reflection. They process the image to provide qualitative or quantitative evaluations of the type, amount and nutritional values of food or beverages. This is a complex task for which a range of automated, semi-automated and manual approaches have been proposed. Standard methods of image segmentation and pattern classification perform poorly in food recognition problems \cite{zhu:2010}. Deep learning algorithms have recently emerged as a powerful image recognition technique \cite{le:2012} . Using vast amounts of data, these algorithms are capable of learning complex representations of images through training to extract the most discriminating features. While these approaches perform better in food recognition than standard methods, they are still in their infancy. The goal is for these algorithms to learn from every image they see and continually improve their accuracy across all food categories. For now, however, automated food recognition and evaluation is not sufficiently accurate for real world use.

Thus, \textit{I Ate This} offers expert evaluation of food photos and coaching. By using technology, these services replace, and are cheaper than face-to-face consultation. Studies show that this kind of remote but personal counselling is a critical component of successful web-based health management programs \cite{tate:2003}.

\begin{marginfigure}[-17pc]
  \begin{minipage}{\marginparwidth}
    \centering
    \includegraphics[width=0.9\marginparwidth]{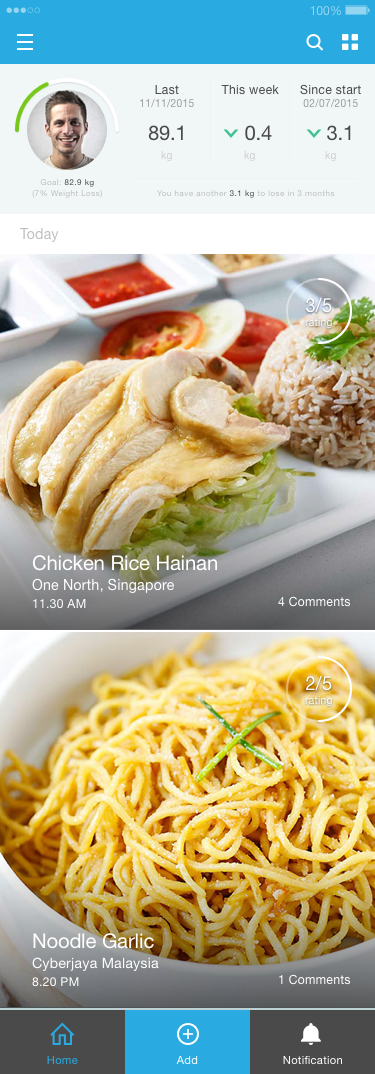}
    \caption{Timeline view showing food photo logs}~\label{fig:timeline}
  \end{minipage}
\end{marginfigure}

\section{System Description}
\begin{figure}[t]
	\centering\includegraphics[width=1.1\columnwidth, height=7.5cm]{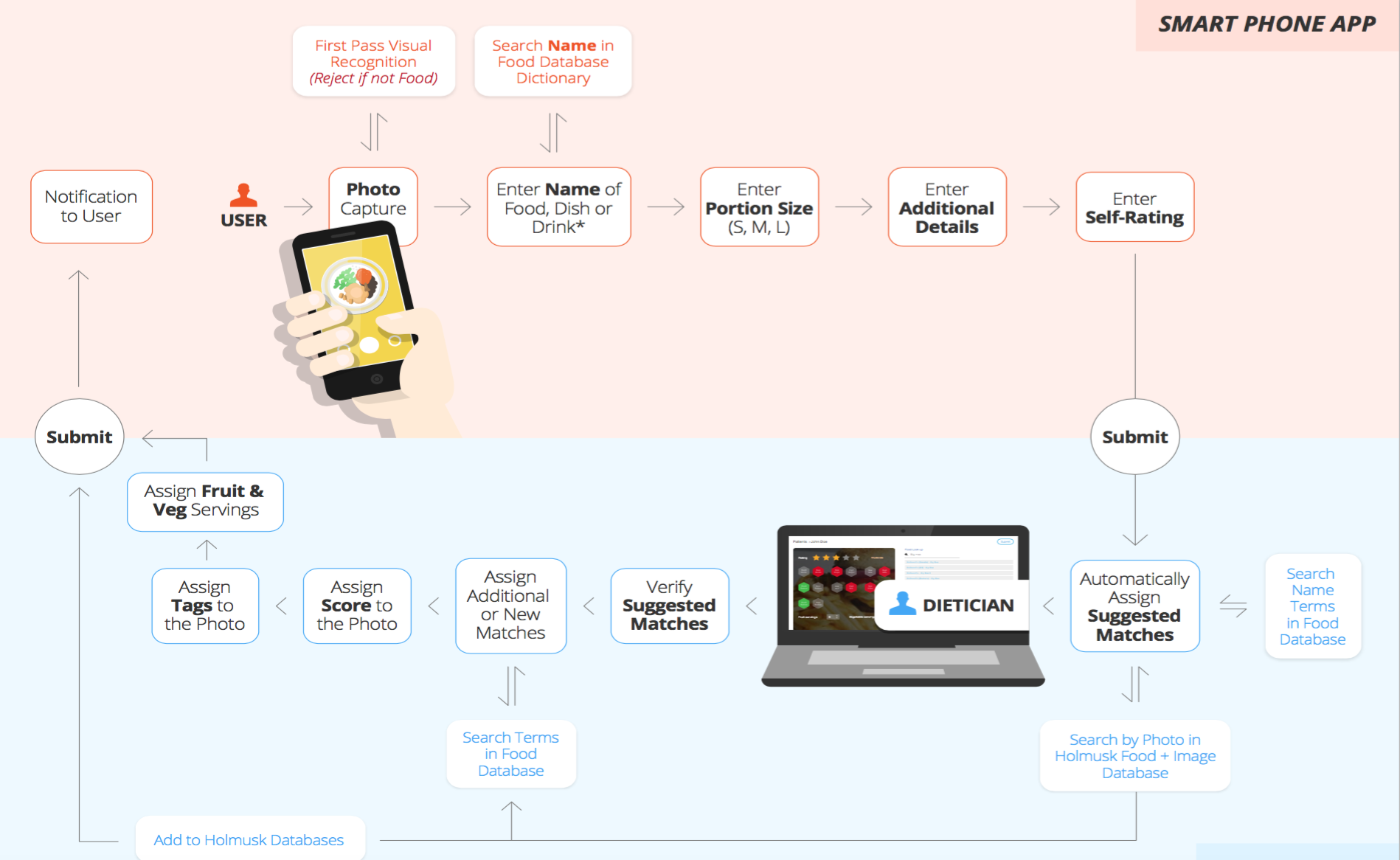}
	\caption{\textit{I Ate This} system flow.}~\label{fig:archi}
\end{figure}

\textit{I Ate This} is a photo-based food journaling system invented by Holmusk \footnote{http://www.holmusk.com}. The system was devised through extensive literature search, usability studies, and consultation with dietitians, behavioural scientists, health educators and UX experts. \textit{I Ate This} encompasses both a smartphone front-end for users to capture and review food photos, and a back office system for dietitians to evaluate and  provide feedback to users based on those photos. The system is built with APIs  that can easily be integrated into other web and mobile apps to support food journaling and feedback. The basic flow of the system is shown in Figure \ref{fig:archi}. 

\subsection{Lightweight Capture}
The \textit{I Ate This} interface for capturing food photos builds on previous research demonstrating the value of a lightweight photo-based interface to support food-related goals. After taking the photo, the user is required to enter a basic text description naming any food, drink or dish in the photo. While the user can enter any text, the task is made simpler through tag suggestions that appear as the user types. Additional optional inputs include: Qualitative description of the portion size (small, medium or large); time and location meta-data for the photo; user self-rating of the food or drink on a 5-point scale (Very Unhealthy to Very Healthy).

\subsection{Expert Evaluation}
\textit{The I Ate This} back office system is designed to assist dietitians in evaluating photos and providing feedback to users. A dietitian can log into the system to view a photostream of `pending' and `scored' submissions. Clicking on any photo brings up the evaluation screen for that submission. It also shows the other submissions by that user, organised by date. The evaluation interface breaks down the task of evaluating photos into three major steps:

\begin{marginfigure}[0pc]
  \begin{minipage}{\marginparwidth}
    \centering
    \includegraphics[width=0.9\marginparwidth]{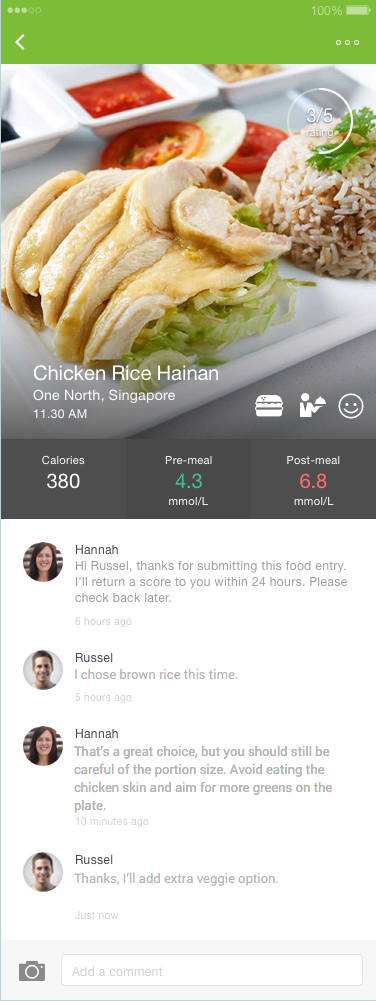}
    \caption{Food detail card showing dietitian score and chat exchange}~\label{fig:fooddetail}
  \end{minipage}
\end{marginfigure}

\begin{enumerate}
	\item \textbf{Match}: In this step the dietitian has to match the photo to one or more food or beverage entries in Holmusk's nutrition database and estimate portions for those matches.
	\item \textbf{Tag}: This step allows the dietitian to tag the photo with traffic light symbols to indicate: Good Carb/Poor Carb, Good Protein/Poor Protein, High Sugar/Low Sugar etc. In addition to these tags, an estimate of the number of Fruit and Vegetable servings is also assigned to the photo.
	\item \textbf{Score}: The final step is to score the photo on a 5-point scale from Very Unhealthy to Very Healthy. This scale is equivalent to the one for user self-rating. A dietitian uses all available information, in addition to their knowledge and experience, to assign this score.
\end{enumerate}

\begin{figure}[t]
	\centering\includegraphics[width=1.1\columnwidth, height=7.5cm]{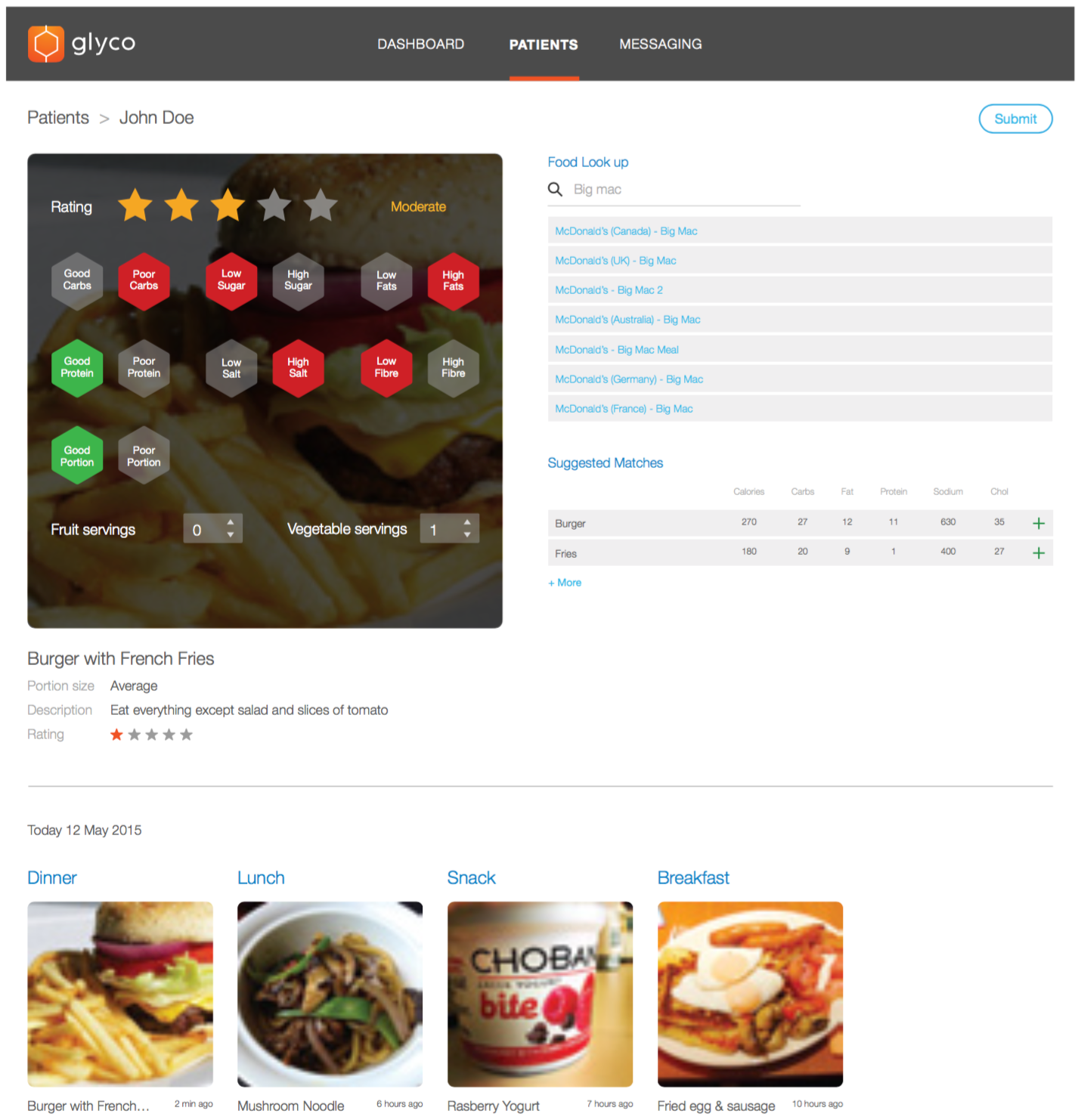}
	\caption{\textit{I Ate This} back office tagging and scoring interface for dietitians.}~\label{fig:backoffice}
\end{figure}

\subsection{Feedback}
\textit{The I Ate This} system is unique in supporting two levels of feedback:
\begin{enumerate}
	\item Quantitative feedback- The database matches assigned to photos by dietitians allows nutrition information to be aggregated and analysed algorithmically to provide automated alerts and insights. For example, a user can be alerted if intake of salt, sugar or calories in a given day exceeds recommendations.
	
	\item Qualitative feedback- \textit{I Ate This} emphasises personalised feedback and encouragement and not just detailed analytics. This is delivered through easy-to-understand tags and scores, and in-app message exchanges with the dietitan. Much of this feedback is linked to individual dietitians who have profiles (complete with photos) on the system.
	\item Automated feedback- \textit{I Ate This} can be integrated with data from Continuous Glucose Monitoring (CGM) devices such as Dexcom, Medtronic CGM, Abbott's and FreeStyle Libre. Logged meal or drink photos are then tagged according to their precise effect on blood glucose levels. For example, foods that lead to out of range glucose values are overlaid with a red marker. Those that have little effect on blood glucose, leading to stable, in-range values are overlaid with a green marker. This offers a very simple and visual way for people with diabetes to use information from their past meals to more accurately manage those same foods in the future.
\end{enumerate}

\section{Conclusion}
By bringing the Remote Food Photography method to a wider audience, \textit{The I Ate This} addresses an important gap in DIY and personal health informatics: a convenient, visual and meaningful way to log food. The feedback component is highly flexible and can support different models e.g. crowd sourced evaluation. In some applications, fully automated evaluations are also possible. For example, we are currently integrating \textit{The I Ate This} with Continuous Glucose Monitoring (CGM) devices to colour code people's food photos according to their effects on blood levels (in-range or out of range). We're also building a database of restaurant meals that would allow them to be logged easily without needing to take a photograph. The user can be guided to relevant menu items based on location. This approach offers huge potential for a community of users to learn about the impact of food choices. 

Ultimately, our goals is to validate the \textit{The I Ate This} system in helping diabetes patients to better manage their condition. 

\bibliographystyle{SIGCHI-Reference-Format}
\bibliography{sample}

\end{document}